\documentclass[aps,prl,reprint,superscriptaddress]{revtex4-2}
\usepackage{graphicx}
\usepackage{braket}
\usepackage{amsmath}
\usepackage{amsthm}
\usepackage{amssymb}
\usepackage[colorlinks,
            linkcolor=blue,
            anchorcolor=black,
            citecolor=blue,
			urlcolor=blue]{hyperref}

\usepackage{algorithm}  
\usepackage{algorithmic}  
\usepackage{makecell}
\usepackage{array}
\usepackage{color}



\begin{document}


\title{Hypergraph product code with 0.2 constant coding rate and high code capacity noise threshold}


\author{Zhengzhong Yi}
\author{Zhipeng Liang} 
\author{Jiahan Chen}
\author{Zicheng Wang}
\author{Xuan Wang}
\email[]{wangxuan@cs.hitsz.edu.cn}
\affiliation{Harbin Institute of Technology, Shenzhen. Shenzhen, 518055, China}


\date{\today}

\begin{abstract}
	The low coding rate of quantum stabilizer codes results in formidable physical qubit overhead when realizing quantum error correcting in engineering. In this letter, we propose a new class of hypergraph-product code called TGRE-hypergraph-product code. This code has constant coding rate 0.2, which is the highest constant coding rate of quantum stabilizer codes to our best knowledge. We perform simulations to test the error correcting capability TGRE-hypergraph-product code and find their code capacity noise threshold in depolarizing noise channel is around 0.096.

\end{abstract}


\maketitle

\section{Introduction}
\label{introduction}
The low coding rate of quantum stabilizer codes has become a major obstacle on the road to realizing quantum computer with enough and sufficiently reliable logical qubits based on quantum stabilizer codes. There is few class of quantum stabilizer codes have constant coding rate, while most classes have zero asymptotic coding rate. Surface code\cite{bravyi1998quantum,fowler2012surface}, XZZX surface code\cite{bonilla2021xzzx}, which are the stabilizer codes closet to engineering application at present, and many other famous quantum stabilizer codes such concatenated code\cite{knill2005quantum}, 3D toric code\cite{kubica2019cellular}, and 4D toric code\cite{breuckmann2016local}, all have zero asymptotic coding rate. (3,4) hypergraph product code\cite{grospellier2021combining}, (5,6) hypergraph product code\cite{grospellier2018numerical} and  (4,5)-hyperbolic
surface code\cite{breuckmann2016constructions} are reported to have constant coding rate. However, their coding rates are 0.04, 0.016 and 0.1 respectively, which are still low compared to classical error correcting codes, such as polar code\cite{arikan2009channel} and LDPC code\cite{ryan2004introduction}.

Hypergraph product codes proposed by Tillich and Zémor\cite{tillich2013quantum} are a class of quantum LDPC codes. Hypergraph product codes are constructed by two binary linear codes. If one chooses the coding rate of these two binary linear codes appropriately, the hypergraph product code constructed by them might have non-zero asymptotic rate.

In this letter, we propose a new class of hypergraph product code named TGRE-hypergraph-product code, whose code distance is $O(\log \sqrt{N})$ and coding rate is constant 0.2, where $N$ is the code length. To our best knowledge, this coding rate is the highest among the existing quantum stabilizer codes. The name TGRE-hypergraph-product code comes from the fact that this code are constructed by two TGRE code\cite{yi2022quantum,yi2024xztype} -- X-TGRE code and Z-TGRE code, whose coding rate are both constant 0.5 and code distance are both $O(\log N)$. 

To test the error correcting capability of TGRE-hypergraph-product code, we perform simulations with the fully-decoupled belief propagation (FDBP) decoding algorithm\cite{yi2023improved}. The simulations show that the code capacity noise threshold\cite{landahl2011fault} of TGRE-hypergraph-product code is around 0.096 in depolarizing noise channel, which is almost twice higher than the code capacity noise threshold of 4D-hyperbolic code\cite{breuckmann2021single}, whose coding rate is the constant 0.18, and almost four times higher than the code capacity noise threshold of (4,5)-hyperbolic surface code\cite{breuckmann2016constructions}. Since the decoding accuracy will be reduced by the increase of the weight of stabilizers, and the weight of stabilizers will increase with the growth of the code length of TGRE-hypergraph-product code, we have reason to believe that the actual code capacity noise threshold should be higher.

\section {Z-TGRE code and X-TGRE code}
\label{Z-TGRE code and X-TGRE code}
\begin{figure}
	
	\centering
	\includegraphics[width=0.5\textwidth]{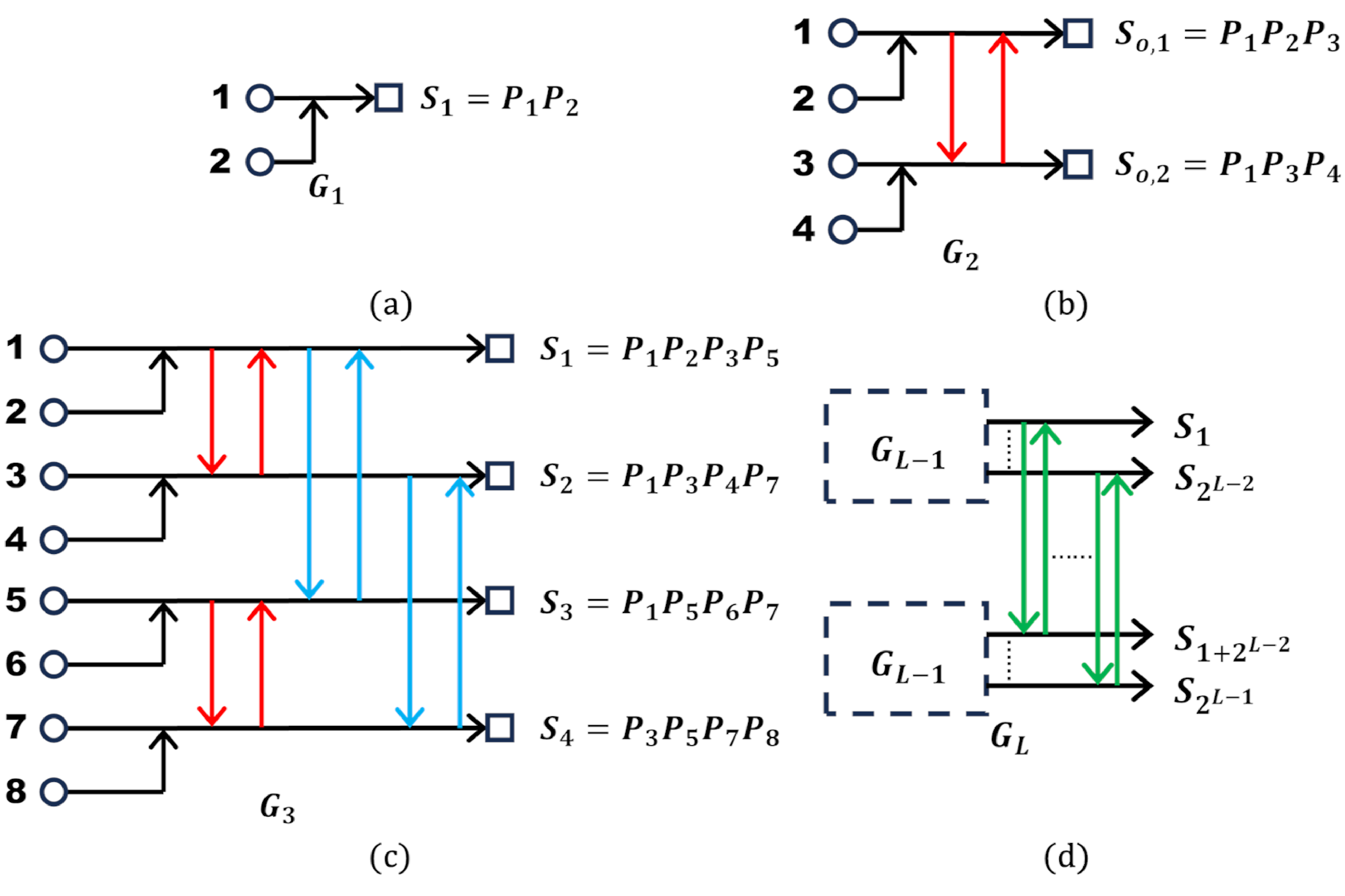}
	\caption{The recursive expansion of Tanner graph of Z-TGRE code and X-TGRE code. The arrow means the corresponding variable node (the qubit) it starts from will join in the corresponding check node (the stabilizer) it ends with. The variable nodes are numbered from 1 to $N=2^L$. $S_i$ represents the $i$th stabilizer generated by the corresponding check nodes in the Tanner graphs. $P \in \{X, Z\}$. For Z-TGRE code, $P=Z$. For X-TGRE, $P=X$. (a) The primal Tanner graph $G_1$ used to recursive expansion. (b) The expanded Tanner graph $G_2$ by the recursive expansion of two primal Tanner graphs $G_1$. (c) The expanded Tanner graph $G_3$ by the recursive expansion of two $G_2$. (d) The expanded Tanner graph Gn by recursive expansion of two $G_{L-1}$.}
	\label{TGRETannergraph}
\end{figure}

The construction of Z-TGRE code and X-TGRE code have been described in detail in \cite{yi2022quantum,yi2024xztype}. Here, we only give out the Tanner graph expansion of them, which is shown in Fig. \ref{TGRETannergraph}. The coding rate of Z-TGRE code and X-TGRE code are both constant 0.5, and the code distance of them are both $O(\log N)$. However, Z-TGRE code can only correct Pauli $X$ and $Y$ errors, while X-TGRE code can only correct Pauli $Z$ and $Y$ errors.

\section {TGRE-hypergraph-product code}
\label{TGRE-hypergraph-product code}

\begin{figure}
	
	\centering
	\includegraphics[width=0.4\textwidth]{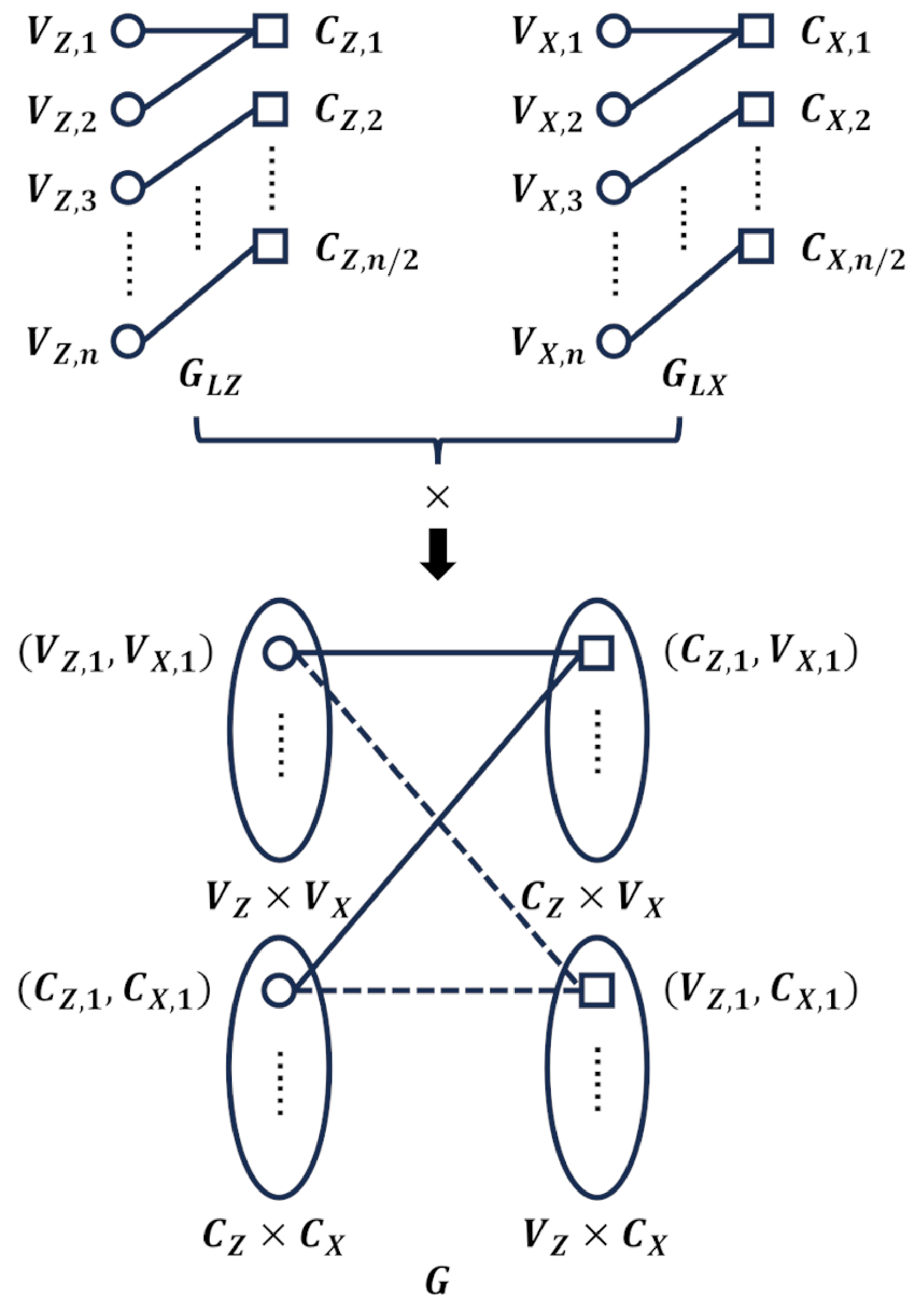}
	\caption{The Tanner graph of TGRE-hypergraph-product code. The "$\times$" means Cartesian product of two sets. $G_{LZ}$ and $G_{LX}$ are the Tanner graphs of the Z-TGRE code and X-TGRE code with code length $n=2^L$, respectively. $G$ is the Tanner graph of the TGRE-hypergraph-product code with code length $N = \frac{5}{4} n^2$.}
	\label{graph product}
\end{figure}

The TGRE-hypergraph-product code with code length $N = \frac{5}{4} n^2$ are constructed by a Z-TGRE code with code length $n$ and a X-TGRE code with code length $n$, in the way shown in Fig. \ref{graph product}. According to Sect. IV in \cite{tillich2013quantum}, the TGRE-hypergraph-product code whose Tanner graph is shown in Fig. \ref{graph product} is a $[\frac{5}{4} n^2, \frac{1}{4}n^2, d]$ code, where $d$ is the code distance of the Z-TGRE code (and also the code distance of the X-TGRE code). Hence, with the growth of the code length of the Z-TGRE code and X-TGRE code used to construct TGRE-hypergraph-product code, the coding rate is always 0.2, and the code distance is  $O(\log n) = O(\log \sqrt{N})$. Besides, according to graph $G$ in Fig. 2, the weight of the stabilizers will increase with the growth of code length at a rate of $O(\sqrt{N})$, or more accurately, $\frac{3}{\sqrt{5}}\sqrt{N}$.

As for the logical operators, they can be given out through the standard form of the check matrix\cite{nielsen2001quantum}.

\section{Simulation results}
\label{simulation results}

\begin{figure}
	
	\centering
	\includegraphics[width=0.5\textwidth]{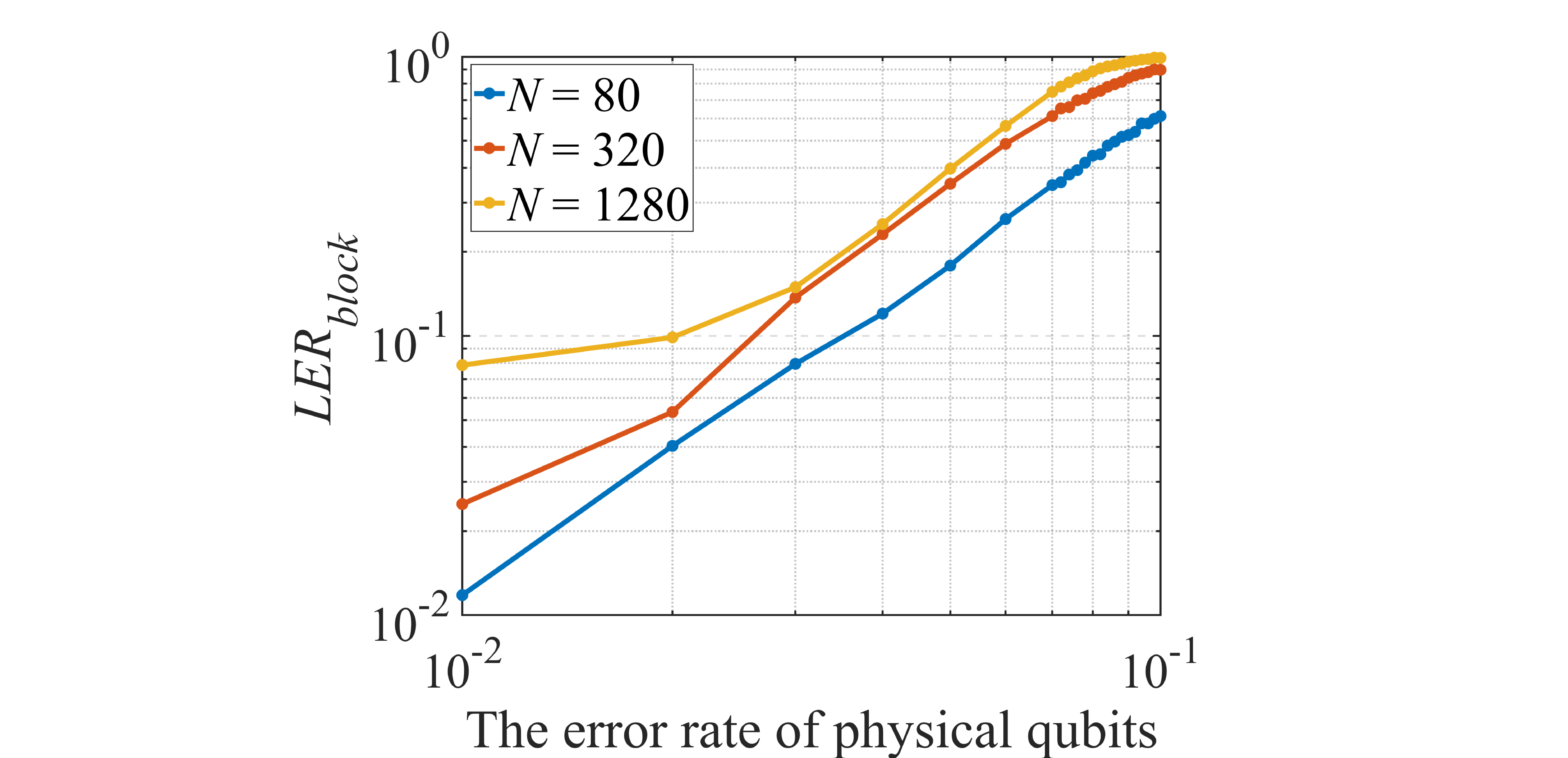}
	\caption{The $LER_{block}$ in depolarizing noise channel.}
	\label{ler_block}
\end{figure}

\begin{figure}
	\centering
	\includegraphics[width=0.5\textwidth]{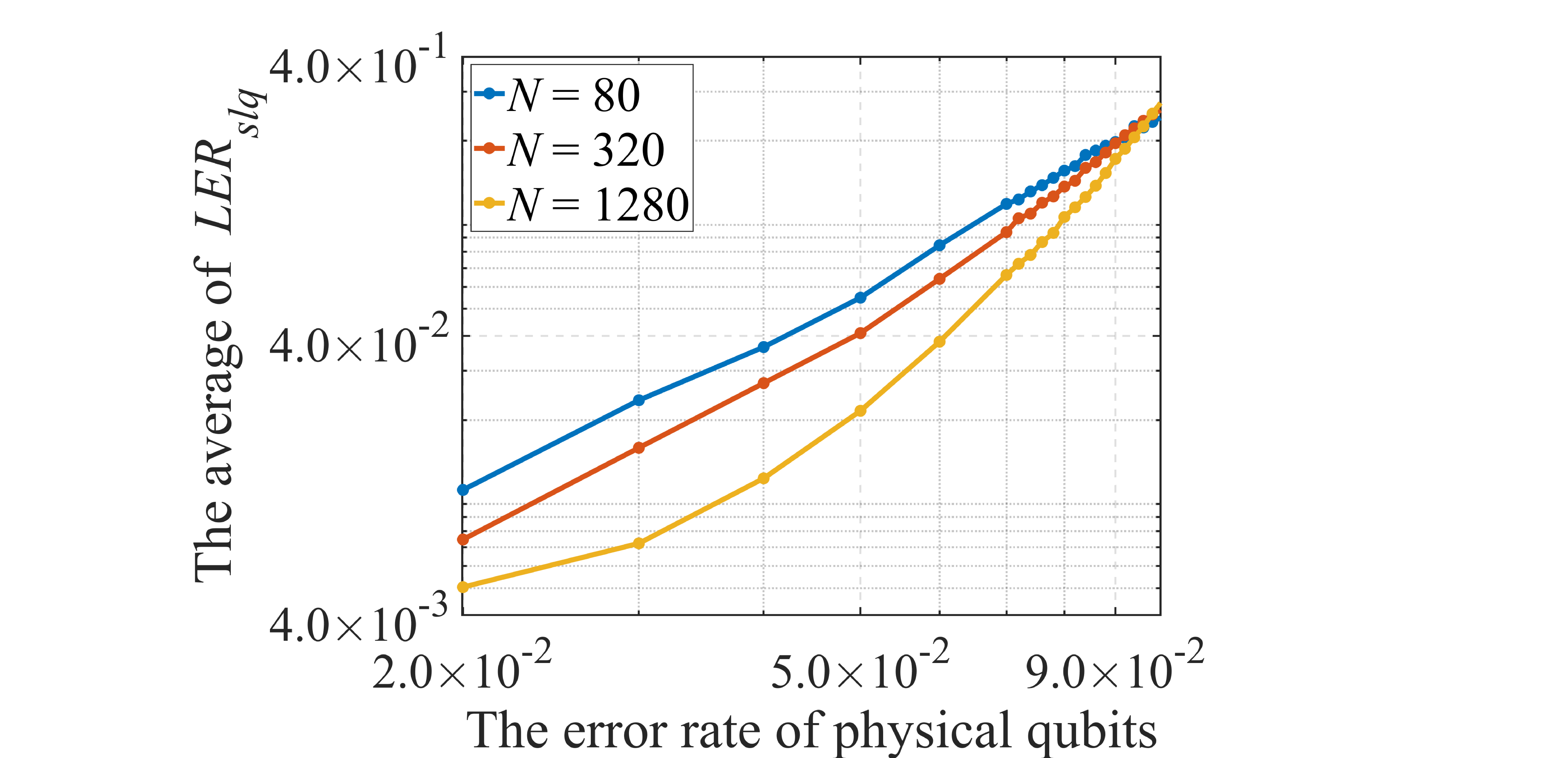}
	\caption{The average of $LER_{slq}$ in depolarizing noise channel.}
	\label{ler_slq}
\end{figure}

\begin{figure}
	\centering
	\includegraphics[width=0.5\textwidth]{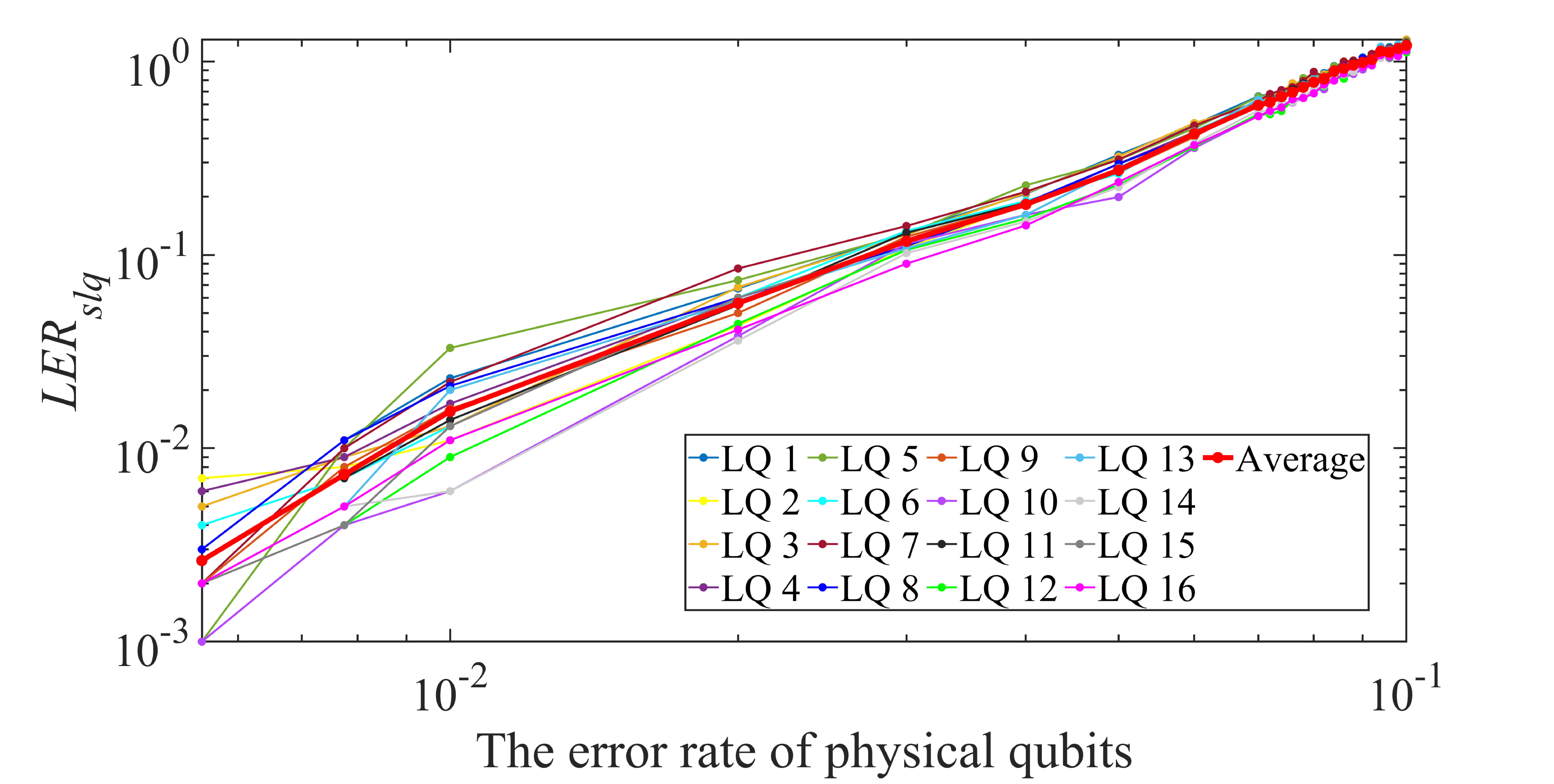}
	\caption{The $LER_{slq}$ every logical qubit when code length $N=80$ in depolarizing noise channel. LQ means logical qubit}
	\label{ler_slq of every logical qubit}
\end{figure}

To test the error correcting capability of TGRE-hypergraph-product code, simulations are performed in depolarizing noise channel with FDBP decoding algorithm. The results are shown in Fig. \ref{ler_block} to Fig. \ref{ler_slq of every logical qubit}. The logical error rate of the whole code block is denoted by $LER_{block}$, and the logical error rate of single logical qubit is denoted by $LER_{slq}$. Fig. \ref{ler_slq} shows the average of $LER_{slq}$ of all logical qubits. Fig. \ref{ler_slq of every logical qubit} shows the $LER_{slq}$ of each logical qubit when the code length is 80.

According to Fig. \ref{ler_slq}, the code capacity noise threshold is around 0.096. It should be noticed that the decoding accuracy of FDBP decoding algorithm will be influenced by the number short loops in the Tanner graph, which is determined by the weight of the stabilizers. The larger the weight is, the more short loops might exist, and the lower the decoding accuracy will be. Hence, we have reason to believe that the actual code capacity noise threshold should be higher than 0.096.

\section {Conclusion}
\label{conclusion}

We use Z-TGRE code and X-TGRE code to construct TGRE-hypergraph-product code by the method in \cite{tillich2013quantum}. The code distance of TGRE-hypergraph-product code is $O(\log \sqrt{N})$ and the coding rate is constant 0.2 which is the highest among the existing quantum stabilizer codes to our best knowledge. In depolarizing channel, with FDBP decoding algorithm, we find the code capacity noise threshold is around 0.096. However, the biggest challenge to apply TGRE-hypergraph-product code in engineering is that the weight of their stabilizers will increase with the growth of code length at a rate of $O(\sqrt{N})$, which do harm to maintaining the locality of operations on the physical qubits. Even so, our work shows that the code construction method of hypergraph-product codes do have the potential to construct quantum stabilizer codes with good performance.

\section*{End Notes}
\subsection*{Acknowledgements}
 This work was supported by the Colleges and Universities Stable Support Project of Shenzhen, China (No.GXWD20220817164856008), the Colleges and Universities Stable Support Project of Shenzhen, China (No.GXWD20220811170225001) and Harbin Institute of Technology, Shenzhen - SpinQ quantum information Joint Research Center Project (No.HITSZ20230111).

\section*{Data Availability}

\bibliography{reference}

%

\end{document}